\documentclass[a4paper]{amsart}
\usepackage{amsmath}%
\usepackage{amsfonts}%
\usepackage{amssymb}%
\usepackage{graphicx}
\theoremstyle{plain}

\newtheorem{conclusion}{Conclusion}

\newtheorem{definition}{Definition}

\numberwithin{equation}{section}
\begin{document}
\title[On numerical stability of recursive PV computation method]{On numerical stability of recursive present value computation method}
\author{Argyn Kuketayev}
\email[Argyn Kuketayev]{jawabean@gwu.edu}%
\date{November 10, 2006}
\subjclass{Primary 65G50, Secondary 65-04 } %
\keywords{present value pv recursive recurrent numerical instability}%

\begin{abstract}
We analyze numerical stability of a recursive computation scheme of present value (PV) amd show that
the absolute error increases exponentially for positive discount rates. 
We show that reversing the direction of calculations in the recurrence equation yields a robust PV computation routine.
\end{abstract}
\maketitle

\section{Introduction}

The concept of time value of money is so important in finance (see chapter 7 in \cite{fabozzi}), that it is not surprising that the  present value (PV) computations are very common in finance and accounting. This work focuses on one of the definitions of PV, which is similar to equation (7-7) in chapter 7 of \cite{fabozzi}.

\begin{definition}
 Present Value (PV) of future cashflows is defined as follows:
\begin{equation}
	PV(r) = \sum^{N}_{m=1}\frac{C_{m}}{(1+r)^m}
	\label{def_PV}
\end{equation}
where $C_{m}$ is the cashflow amount in month $m$, $N$ is the month number of last non-zero cashflow, $r$ is a monthly 
discount rate, or required rate of return.
Cash outflows are negative, and cash inflows are positive.
\end{definition}

\begin{definition}
 Net Present Value (NPV) of cashflows is defined as follows :
\begin{equation}
	NPV(r) =  C_0 + PV(r)
\end{equation}
where an initial cash outflow is $C_{0}$, i.e. month zero. 
Cash outflows are negative, and cash inflows are positive.
\end{definition}

Of course, the most trivial method of calculation of PV is by direct summation of discounted cashflows $\frac{C_{m}}{(1+r)^m}$, i.e.  
just like in the definition. However, it is easy to to derive a recurrence equation. First, let us define a  present value of the 
subset of a given cashflow as follows.
\begin{definition}
The partial PV of a cashflow is defined by equation:
\begin{equation}
	PV_k = \sum^{N}_{m=k+1}\frac{C_{m}}{(1+r)^m}
	\label{ppv}
\end{equation}
\end{definition}

Obviously, $PV_0$ is PV of an entire cashflow excluding initial payment, i.e. $PV_0\equiv PV(r)$. 
From the time value point of view, $PV_k$ is nothing but the 
present value of the remaining cashflows at month $k$ (exclusive). Knowing a partial PV in previous month, it is easy to calculate it 
for the current month using the recurrence equation:
\begin{equation}
	 PV_k= (1+r)\cdot PV_{k-1}-C_{k}
	\label{recur_pv}
\end{equation}

This formula can easily be derived as follows:
\[
PV_k = \sum^{N}_{m=k+1}\frac{C_{m}}{(1+r)^m}= (1+r)\sum^{N}_{m=k}\frac{C_{m}}{(1+r)^m}-C_{k}
\]

\section{Recurrent PV computation use case}

We believe that the equation \ref{recur_pv} is used in practice. Consider the Table 1 in Appendix B of \cite{FAS91}. 
The document was published by Financial Accounting Standards Board (http://www.fasb.org/) and it sets a
guideline on amortization of certain fees and costs associated with loan origination. 
We refer to chapter 15 and page 688 in \cite{fabozzi} for 
discussion of \emph{amortization}, \emph{amortization schedule} and \emph{loan origination fees}. 
Table \ref{tbl_fas91} is essentially a fragment of the above mentioned table with insignificant modifications. 
We used Google Spreadsheet application \cite{google} to create this table by implementing
FAS 91 rules in the worksheet.
We show it here in order to make a point that the recursive PV computation algorithm is actually used in practice and also 
to explain why it is needed.

\begin{table}[ht]
	\centering
	\caption{Amortization Based on Contractual Payment Term}
	\label{tbl_fas91}
		\begin{tabular}{llllll}
&(1)&(2)&(3)&(4)&(7)\\
Year&Cash (Out)&Stated&Amortization&Interest&Carrying\\
&Inflow&Interest&&Income&Amount\\
&-\$98,000.00&&&&98,000.00\\
1&\$16,274.54&10,000.00&263.52&10,263.52&91,988.98\\
2&\$16,274.54&9,372.55&261.44&9,633.99&85,348.43\\
3&\$16,274.54&8,682.35&256.18&8,938.53&78,012.42\\
4&\$16,274.54&7,923.13&247.10&8,170.23&69,908.11\\
5&\$16,274.54&7,087.99&233.48&7,321.46&60,955.04\\
6&\$16,274.54&6,169.33&214.48&6,383.81&51,064.31\\
7&\$16,274.54&5,158.81&189.15&5,347.96&40,137.72\\
8&\$16,274.54&4,047.24&156.38&4,203.62&28,066.80\\
9&\$16,274.54&2,824.51&114.92&2,939.43&14,731.69\\
10&\$16,274.54&1,479.50&63.34&1,542.85&0.00\\

&Total &&\$2,000.00	\\
&amortization\\
		\end{tabular}
\end{table}

We have to introduce the concept of the internal rate of return (IRR) (see page 202 in \cite{fabozzi}) in order to deconstruct 
Table \ref{tbl_fas91}.

\begin{definition}
Internal rate of return (IRR) of cashflows is a rate $r$, which makes NPV(r) equal to zero and is defined by the following equation:
\begin{equation}
	NPV(irr) = \sum^{N}_{m=0}\frac{C_{m}}{(1+irr)^m}\equiv 0
	\label{irr}
\end{equation}
\end{definition}

In the Table \ref{tbl_fas91} column (1) contains cashflow amounts for each year. In FAS 91 examples years are used instead of months. It does not
change anything in principle if we assume that years are months, for the time being. In our notation $C_0=-98,000.00$ and $N=10$. 
Next, the IRR is calculated for this cashflow and it is approximately $10.47\%$. Now, column (7) (carrying amount) is actually 
$PV_{k}(irr)$, e.g. carrying amount in year 9: \[NPV_9(irr)=\frac{C_{10}}{1+irr}=\frac{16,274.54}{1+0.1047}=14,731.69\]

The following is an excerpt from the description to Table 1 in \cite{FAS91}:
\begin{quotation}
Computations:	\\		
Column (1): Contractual payments\\			
Column (2): Column (5) for prior year $\times$  the loan's stated interest rate (10\%)\\
Column (3): Column (4) - Column (2)			\\
Column (4): Column (7) for prior year $\times$ the effective interest rate $(10.4736\%)^b$ 			\\
Column (5): Column (5) for prior year - (Column (1) - Column (2))			\\
Column (6): Initial net fees - amortization to date			\\
Column (7): Column (5) - Column (6)			\\

\end{quotation}

In this description IRR is referred to as \emph{effective interest rate}. Also, in our copy of this table we omitted columns (5) and (6)
in order to save the space. We leave it to the reader to prove that carrying amount is equivalent to PV (definition \ref{ppv}),
and that it is computed recursively (equation \ref{recur_pv}) starting from the top of the table. We shall call this algorithm \emph{forward} recursive 
PV computation scheme, as opposed to \emph{backwards} scheme, which will be described later. The table itself is
often called \emph{amortization schedule}. 

Why not compute carrying amounts directly using PV formula \ref{def_PV}? Consider the following use case. A company may have a huge portfolio 
of loans, maybe 500,000 loans. Long-term loans, such as mortgages, may have scheduled terms of 30 years (360 months) or even more. 
Often more than one fee is amortized. Direct calculation of NPV
for each month in the schedule yields excessive redundant summation of the same discounted cashflow amounts over and over.
Therefore, it is very desirable to have a recursive PV computation method at hand.

\section{Numerical instability of forward recursive PV computation for positive IRR}

Let us analyze absolute error accumulation when using \emph{forward} PV computation(eq. \ref{recur_pv}). The following equations describe accumulation
of rounding errors in floating-point addition/subtraction and multiplication.
\begin{equation}
	\Delta_{a\oplus b} \approx \Delta_{a} + \Delta_{b}
\end{equation}
\begin{equation}
	\Delta_{a\otimes b} \approx \Delta_{a}\cdot |b|+ \Delta_{b}\cdot |a|
\end{equation}

These equations are commonly used in analysis of error propagation. Essentially, they are equivalent to the notion of accumulation of absolute errors
in summation/subtraction operations, and accumulation of relative errors in multiplication/division operations.
One can derive them using partial derivatives as follows (see 2.11 in \cite{hamming})
\begin{equation}
	\Delta_{a\otimes b} \approx |\frac{\partial}{\partial a}(a\cdot b)|\cdot\Delta_{a} + |\frac{\partial}{\partial b}(a\cdot b)|\cdot\Delta_{b}
	\label{err_prop}
\end{equation}

Applying these equations to equation \ref{recur_pv}, we can deduct the following 
\begin{equation}
	\Delta_{PV_k} \approx |1+r| \cdot \Delta_{PV_{k-1}} + |PV_{k-1}| \cdot \Delta_{r} + \Delta_{C_k}
\end{equation}

We shall assume that $\Delta_{C_k} \equiv0$, i.e. cashflow amounts are precisely accurate. Also, $PV_{0} \equiv C_0$, because 
we used $irr$ as a discount rate. Therefore, $\Delta_{PV_{0}}\equiv 0$ and the following is true
\begin{equation}
	\Delta_{PV_1} \approx PV_{0} \cdot \Delta_{r} \equiv |C_0|\cdot \Delta_{r}
\end{equation}
\begin{equation}
	\Delta_{PV_2} \approx |1+r| \cdot |C_0|\cdot \Delta_{r} + |PV_{k-1}| \cdot \Delta_{r} 
\end{equation}
\begin{equation}
	\Delta_{PV_3} \approx |1+r|^2 \cdot |C_0|\cdot \Delta_{r} + |1+r| \cdot |PV_{k-1}| \cdot \Delta_{r}  + |PV_{k-1}| \cdot \Delta_{r} 
\end{equation} and so on. It is easy to show by simple mathematical induction that the absolute error of present value has a term, 
which increases exponentially with $k$ for positive rates $r$:
\begin{equation}
\Delta_{PV_k} \approx |1+r|^{k-1} \cdot |C_0|\cdot \Delta_{r}
\label{fwd_err}
\end{equation} 
Note that $\Delta_{r}$ can not be zero, because it is computed by numerically solving equation \ref{irr}. Therefore, we state
that the forward recursive PV computation is numerically unstable for $r>0$.

In order to demonstrate our finding, let us compute $PV_k$ for every year in Table \ref{tbl_fas91} using direct summation formula \ref{ppv},
then compute the rate of year-on-year increase of the difference between this value and the carrying amount. The results are shown in Table \ref{tbl_fas91_err}\footnote{Note that the discount rate in this sample is 10.47\%.}.  If we assume that direct $PV_k$ computation by equation \ref{ppv} is a better estimate of true present value, then 
column (9) confirms the correctness of our error analysis as per equation \ref{fwd_err}.

\begin{table}[ht]
	\centering
	\caption{Amortization Error Accumulation Exhibit}
	\label{tbl_fas91_err}
		\begin{tabular}{r|l|l}
&(8)&(9)\\
\hline
Year&$PV_k$ (as in \ref{ppv})& \% increase from\\
& - Column (7)&previous year\\
\hline
1&0.000000000509317	\\
2&0.000000000567525&11.43\%	\\
3&0.000000000640284&12.82\%	\\
4&0.000000000698492&9.09\%	\\
5&0.000000000778527&11.46\%	\\
6&0.000000000858563&10.28\%	\\
7&0.000000000945874&10.17\%	\\
8&0.0000000010441	&10.38\%	\\
9&0.000000001149601	&10.10\%\\
10&0.000000001269882	&10.46\%\\
		\end{tabular}
\end{table}

\section{Backward recursive PV computation is robust for positive IRR}

Let us look at equation \ref{recur_pv} and its absolute error formula \ref{fwd_err}. There seem to be two problems.
\begin{itemize}
	\item Every next present value is being multiplied by factor $1+r$, so the absolute error of $PV_k$ should 
increase every month by percentage equal to $r$ for positive rates.
	\item  $PV_k$ itself is decreasing toward the end of the schedule
until it finally reaches zero. 

\end{itemize}
Combined, these two factors should magnify the relative error of the present at every step of the calculations.

A heuristic suggests considering to reverse the recurrence equation \ref{recur_pv}

\begin{equation}
	 PV_{k-1}= \frac{1}{1+r} (PV_k +C_{k})
	\label{bwd_pv}
\end{equation}

We shall call the new recurrence equation a \emph{backward} recursive computation method. It seems to address both mentioned issues.
\begin{itemize}
	\item Its next step is evaluated by division of the previous value by $1+r$, subsequently the absolute error of the present values should not
be magnified.
	\item   The sum in the second factor generally increases on every step, i.e. unlike \ref{recur_pv} it does not add to
 relative errors of computed values.
\end{itemize}

Like we did it earlier for \emph{forward} method, let us conduct a simple error propagation analysis. First, we shall define the absolute error of $\frac{1}{1+r}$ using equation \ref{err_prop}

\begin{equation}
	\Delta_{\frac{1}{1+r}} \approx \frac{\Delta_{r}}{(1+r)^2}
\end{equation}

Now, we can estimate the error of the \emph{backward} recursive computation as follows, again, assuming $\Delta_{C_k}\equiv 0$
\begin{equation}
	\Delta_{PV_{k-1}} \approx \frac{\Delta_{PV_{k}}}{|1+r|} + \frac{|PV_{k}|\cdot \Delta_{r}}{(1+r)^2} 
\end{equation}

Note, that 
$PV_N\equiv 0$ and $\Delta_{PV_{N}}\equiv 0$, because we are using IRR as a discount rate.
Now we can find out absolute errors of first steps of \emph{backward} computation. 

\begin{equation}
	\Delta_{PV_{N-1}} \approx \frac{0}{|1+r|} + \frac{0\cdot \Delta_{r}}{(1+r)^2} = 0
\end{equation}

\begin{equation}
	\Delta_{PV_{N-2}} \approx \frac{0}{|1+r|} + \frac{|PV_{N-1}|\cdot \Delta_{r}}{(1+r)^2} = \frac{|PV_{N-1}|\cdot \Delta_{r}}{(1+r)^2}
\end{equation}

\begin{equation}
	\Delta_{PV_{N-3}} \approx \frac{|PV_{N-1}|\cdot \Delta_{r}}{|1+r|^3} + \frac{|PV_{N-2}|\cdot \Delta_{r}}{(1+r)^2} 
\end{equation}
and so on. We can generalize these equations for positive discount rates $r$ using mathematical induction as follows
\begin{equation}
	\Delta_{PV_{k}} \approx  \frac{|PV_{k+1}|\cdot \Delta_{r}}{(1+r)^2} +O \left(\frac{|PV_{k+2}|\cdot \Delta_{r}}{|1+r|^3}\right) 
\end{equation}

As expected, for positive discount rates $r$, there are no exponentially increasing terms. The only increasing factor
of $\Delta_{PV_{k}}$ is of order $PV_{k+1}$. However, it does not magnify the relative error, because we know that generally $PV_k>PV_{k+1}$. 

For negative discount rates $r$, $\Delta_{PV_{N-k}}$ will have terms like $\frac{|PV_{N-m+1}|\cdot \Delta_{r}}{|1+r|^{k-m}}$. 
Similar to \emph{forward} computation with positive $r$, the factor $\frac{1}{|1+r|^{k-m}}$ increases exponentially with $k$.
However, as it was stated earlier, $PV_k$ decreases toward the end of an amortization schedule. Therefore, in \emph{backward} computation 
$PV_{N-m+1}$ in numerator should compensate, at least partially, the shrinking denominator $|1+r|^{k-m}$.

In order to test our findings, we produced tables \ref{tbl_fas91} and \ref{tbl_fas91_err} using \emph{backward} computation scheme. 
As expected, we did not
observe exponential increase of absolute errors in recursively computed present values. 
In order to deal with this instability,
as it is suggested by theoretical error propagation analysis, it is necessary to increase an accuracy of IRR calculation, 
i.e., in other words, decrease
$\Delta_{r}$. However, the exponential character of absolute error accumulation and the finite precision of floating-point arithmetics
limits the remedial power of this approach. 

\begin{conclusion}
We demonstrated numerical instability of forward recursive present value computation method.
We showed that this method is well known and might be used in accounting and financial software.
We stated that backward recursive present value computation method is robust, and suggested
to consider using it instead of forward method for positive discount rates.
\end{conclusion}

\end{document}